\documentclass[twocolumn,secnumarabic,amssymb, amsmath, nofootinbib,tightenlines,
nobibnotes, aps, prl,epsfig]{revtex4}
\usepackage{graphicx}
\usepackage{dcolumn}
\usepackage{bm}
\begin{document}
\preprint{APS/123-QED}
\title{Reduced cross-section in Electron-Ion Colliders at small $x$}

\author{G.R.Boroun}%
 \email{boroun@razi.ac.ir }
 \affiliation{Department of Physics, Razi University, Kermanshah
67149, Iran}%
\author{B.Rezaei }
\altaffiliation{brezaei@razi.ac.ir}
\affiliation{Department of Physics, Razi University, Kermanshah
67149, Iran}
\date{\today}
\begin{abstract}
The nuclear reduced cross section $\sigma^{A}_{r}$, in the
kinematic range of the electron-Ion collider  with center-of-mass
energy $\sqrt{s}=140~\mathrm{GeV}$ and $y{\leq}1$, is discussed.
The importance of the nuclear longitudinal structure function
$F^{A}_{L}$ and its behavior owing to the impact parameter for the
heavy and light nucleus of Pb-208 and C-12 at $Q^2=5$ and
$10~\mathrm{GeV}^2$ is considered. The dependence of the ratios
$R^{A}_{F_{L}}$ and $R^{A}_{\sigma}$  on the impact parameter and
the expanding point of the gluon density at small $x$ are
investigated. The factorized form of parton distributions in
nuclei is used in HIJING2.0 model.\\

\end{abstract}
 \pacs{***}
\maketitle
\subsection{I. Introduction}
The shadowing seen [1] in nuclear deep inelastic scattering (DIS)
at small Bjorken $x$ (where $x$ is the longitudinal momentum
fraction of the nucleon carried by the struck parton) is a
distinct phenomenon when the nuclear structure functions compared
to that for free nucleons. The DIS of leptons off nuclei offers a
unique opportunity for the detailed mapping of nuclear structure
in the wide range of $x$ and $Q^2$ (where $Q^2$ the (negative)
boson virtuality exchanged between the lepton and the quark from
the nucleus) . The shadowing at small $x$ states that the
structure function per nucleon is smaller in nuclei than in a free
nucleon and characterized by depletion of $F_{2}^{A}$ with respect
to $F_{2}^{p}$ [2-4].\\
Nuclear structure functions ($F_{2}^{A}$ and $F_{L}^{A}$) are
needed in the computation of reduced cross section in high energy
nuclear collisions. In the framework of DIS, it is possible to
extract the nuclear structure functions from the singlet and gluon
distribution functions as parameterized in Refs.[5,6]. These
parameterizations describe very well the available experimental
data of the reduced cross section. They are provide a behavior of
the cross sections ${\sim}{\ln}^{2}1/x$, in an agreement with the
Froissart predictions [7]. These results for the longitudinal
structure function have been reported in Refs.[8-11], where the
longitudinal structure function extracted at moderate and low
values of $x$ is in reasonably good agreement with the available
experimental data.\\
The nuclear reduced cross section $\sigma_{r}^{A}$ can be
standardly defined via the structure functions $F_{2}^{A}$ and
$F_{L}^{A}$ for the collision of the transversal and longitudinal
virtual photon of momentum $q$, $q^2=-Q^2$, on the nucleus A by
\begin{eqnarray}
\sigma_{r}^{A}(x,Q^2)=F_{2}^{A}(x,Q^2)-\frac{y^2}{Y_{+}}F_{L}^{A}(x,Q^2),
\end{eqnarray}
where $Y_{+}=1+(1-y)^2$ with the inelasticity variable $y$. The
nuclear structure function of a nuclei A with Z protons and N=A-Z
neutrons is
\begin{eqnarray}
F_{2}^{A}(x,Q^2)=\sum_{q}e^{2}_{q}\bigg{[}xf_{q}^{p/A}(x,Q^2)+xf_{\overline{q}}^{p/A}(x,Q^2)
\bigg{]},
\end{eqnarray}
where the nuclear parton distribution functions (nPDFs) read
\begin{eqnarray}
f_{i}^{A}(x,Q^2)=\frac{Z}{A}f_{i}^{p/A}(x,Q^2)+\frac{N}{A}f_{i}^{n/A}(x,Q^2),
\end{eqnarray}
and $f_{i}^{p/A}$ and $f_{i}^{n/A}$ are the parton distribution
functions (PDFs) of a bound proton and neutron in a nuclei A
respectively. Nuclear effects in the nPDF and in the structure
function $F_{2}^{A}$ are defined by the following forms
\begin{eqnarray}
R_{i}^{A}(x)=\frac{xf_{i}^{A}(x,Q^2)}{Axf_{i}^{p}(x,Q^2)},\nonumber\\
R_{F_{2}}^{A}(x)=\frac{F_{2}^{A}(x,Q^2)}{AF_{2}^{p}(x,Q^2)},
\end{eqnarray}
which is defined by small $x$ shadowing [12]. The different
perturbative QCD-based models at scales relevant to LHC and RHIC
experiments for nuclear shadowing of gluons have been investigated
in Ref.[13]. The shadowing of gluons is defined by
\begin{eqnarray}
R_{G}^{A}(x)=\frac{xg^{A}(x,Q^2)}{Axg^{p}(x,Q^2)}.
\end{eqnarray}
This phenomenon is defined as a recombination effect at small $x$
due to high gluon number density in the Infinite Momentum Frame
(IMF) where nucleus is fast. The behavior of the small $x$ gluons
from different nucleons leads to depletion of the nuclear density,
which is  referred to as saturation of gluon density. This slows
down the unlimited growth of the gluon distribution function. The
shadowing seen in nuclear DIS [14] for $x{\leq}0.01$ is
characterized by comparison of the nuclear structure function with
the proton structure function. Within the color glass condensate
(CGC) saturation approach [15], this effect predicted a saturation
scale $Q_{s}(x)$ which establishes the region where the increasing
of the unintegrated gluon distribution (UGD) on $x$ is tamed. In
the simplest form, the saturation scale is estimate to be
$Q_{s,A}^{2}{\propto}A^{\frac{1}{3}}$. Saturation effect at small
$x$ will be one of the key physics goals of an Electron-Ion
Collider (EIC) [16,17]. This construction, with a possibility to
operate with a wide variety of nuclei, will extend the kinematic
acceptance to $x$ and $Q^2$ with varying center-of-mass energies
from $\sqrt{s}=20-140~\mathrm{GeV}$ and $0.01{\leq}y{\leq}0.95$
which is very important to effectively constrain nPDFs [18,19].
The interactions in the EIC will be probes regions of
progressively higher gluon density in nuclei.\\
 The purpose of this
paper is to evaluate the nuclear reduced cross section in the
kinematic regions corresponding to the EIC for $eA$ collision. We
produce results for the $F_{2}^{A}$ and $F_{L}^{A}$ using the
parametrization method [5,6] owing to the adopted $x$
dependence of the ratio of structure functions.\\
The structure of the manuscript is as follows. In Section II we
review the method to compute the nuclear reduced cross section as
the virtual photon-nucleus cross section is smaller than A times
the photon-nucleon cross section. Results for the parametrization
of the proton structure functions are shown and discussed in
Section III. Finally, we present our
results and conclusions in Section IV.\\

\subsection{II. $\gamma^{*}$-A scattering}

The reduced cross section for electron scattering from a nucleus
directly is dependent on the double differential cross section as
\begin{eqnarray}
\sigma^{A}_{r}{\equiv}\bigg{(}
\frac{d^{2}\sigma^{lA}}{dxdQ^2}\bigg{)}\frac{xQ^4}{2{\pi}{\alpha^{2}_{em}}[1+(1-y)^2]},
\end{eqnarray}
where $\frac{d^{2}\sigma^{lA}}{dxdQ^2}$ is dependent on the lepton
and nucleus tensors [4,12,20]. The deeply inelastic lepton-nucleus
scattering (DIS) experiments are the cleanest way of getting
information of the nPDF. The nuclear tensor $W^{A}_{\mu{\nu}}$ is
defined by the nucleon tensor as
\begin{eqnarray}
W^{A}_{\mu{\nu}}(p_{A},q)=\int{d^{4}p_{N}}S(p_{N})W^{N}_{\mu{\nu}}(p_{N},q),
\end{eqnarray}
where $P_{A,N}$ are the nuclear and nucleon momentums
respectively, and the spectral function $S(p_{N})$ is the nucleon
momentum distribution in a nucleus. The nuclear transverse and
longitudinal structure functions are defined in terms of the
structure functions $W_{1}^{A}$ and $W_{2}^{A}$ where they are in
accordance with the photon polarization vectors\footnote{The
polarization vectors for the virtual photon are defined by
$\epsilon_{\pm}=\mp(0,1,\pm{i},0)/\sqrt{2}$ and
$\epsilon_{0}=\mp(\sqrt{\nu^{2}+Q^2},0,0,\nu)/\sqrt{Q^2}$ where
$\nu$ is the energy transfer $\nu=E_{e}-E'_{e}$.} [21]. In the
nucleus rest frame they are defined by\footnote{For future
discussion, please refer to Ref.[21].}
\begin{eqnarray}
F^{A}_{1}(x_{A},Q^2)&=&\sqrt{p^{2}_{A}}W_{1}^{A}(p_{A},q),\nonumber\\
F^{A}_{2}(x_{A},Q^2)&=&\frac{p_{A}.q}{\sqrt{p^{2}_{A}}}W_{2}^{A}(p_{A},q),
\end{eqnarray}
where $x_{A}=\frac{M_{N}}{M_{A}}x$ and the longitudinal nuclear
structure function reads
\begin{eqnarray}
F^{A}_{L}(x_{A},Q^2)=\bigg{(}1+\frac{Q^2}{\nu^2}\bigg{)}F^{A}_{2}(x_{A},Q^2)
-2x_{A}F^{A}_{1}(x_{A},Q^2).
\end{eqnarray}
The nuclear structure function $F_{2}^A$ is the structure function
$F_{2}$ for a nuclear target A which can be standardly defined via
the cross sections $\sigma_{T,L}$ for the collision of the
transversal (T) or longitudinal (L) virtual photon on the nucleus
A as
\begin{eqnarray}
F_{2}^A(x,Q^2)=\frac{Q^2}{4{\pi^2}\alpha_{em}}(\sigma_{T}^{\gamma^{*}A}+\sigma_{L}^{\gamma^{*}A})
(x,Q^2).
\end{eqnarray}
Over a wide saturation kinematic range (i.e.,
$10^{-5}{\leq}x{\leq}0.1$ and
$0.05{\leq}Q^2{\leq}100~\mathrm{GeV}^2$), reduction of the nuclear
reduced cross section with respect to the $\sigma_{r}^{p}$ is
given by
\begin{eqnarray}
R^{A}_{\sigma}(x){\equiv}\frac{\sigma^{A}_{r}(x,Q^2)}{A\sigma^{A}_{r}(x,Q^2)}=
\frac{F_{2}^{A}(x,Q^2)-\frac{y^2}{Y_{+}}F_{L}^{A}(x,Q^2)}{A\bigg{[}F_{2}^{p}(x,Q^2)-\frac{y^2}{Y_{+}}F_{L}^{p}(x,Q^2)\bigg{]}},~
\end{eqnarray}
where the behavior of the longitudinal structure functions is
dependent on the gluon distribution at small $x$. The longitudinal
structure function according to the Altarelli-Martinelli [22]
equation in QCD at small $x$ using the expansion method [23] for
the gluon distribution function at an arbitrary point $z=a$ has
been obtained at the leading-order (LO) approximation in Ref.[24]
by the following form\footnote{At small $x$ the gluon distribution
is the dominant one in the longitudinal structure function.}
\begin{eqnarray}
F_{L}^{A}(x,Q^2){\simeq}\frac{10\alpha_{s}}{27\pi}xg^{A}(\frac{x}{1-a}(\frac{3}{2}-a),Q^2).
\end{eqnarray}
The normalized ratio of the longitudinal structure function is
defined
\begin{eqnarray}
R^{A}_{F_{L}}(x)=\frac{F_{L}^{A}(x,Q^2)}{AF_{L}^{p}(x,Q^2)}{\simeq}R^{A}_{G}(kx),
\end{eqnarray}
where $k=\frac{3-2a}{2-2a}$. Therefore, the ratio $R^{A}_{\sigma}$
at small $x$ can be approximately expressed in terms of the proton
structure functions as
\begin{eqnarray}
R^{A}_{\sigma}&{=}&
\frac{F_{2}^{A}(x,Q^2)-\frac{y^2}{Y_{+}}F_{L}^{A}(x,Q^2)}{A\bigg{[}F_{2}^{p}(x,Q^2)-\frac{y^2}{Y_{+}}F_{L}^{p}(x,Q^2)\bigg{]}}\\
&&\simeq\frac{R^{A}_{F_{2}}(x)F_{2}^{p}(x,Q^2)-\frac{y^2}{Y_{+}}\frac{10\alpha_{s}}{27\pi}R^{A}_{G}(kx)xg^{p}(kx,Q^2)}
{F_{2}^{p}(x,Q^2)-\frac{y^2}{Y_{+}}\frac{10\alpha_{s}}{27\pi}xg^{p}(kx,Q^2)}.\nonumber
\end{eqnarray}
In the following, we discuss the parametrization for the nuclear
parton distributions into the parametrization of
$F_{2}^{p}(x,Q^2)$, $F_{L}^{p}(x,Q^2)$ and $xg^{p}(x,Q^2)$ [5-11].\\

\subsection{III. Structure Functions}

At small values of the Bjorken variable $x$, the parametrization
of $F_{2}^{p}(x,Q^2)$ is presented based on an accurate fit to the
HERA data for $Q^2{\geq}0.15~\mathrm{GeV}^2$ in Ref.[5]. This
parametrization by the form
\begin{eqnarray}
F^{p}_{ 2}(x,Q^{2})& =& D(Q^{2})(1-
x)^{n}\sum_{m=0}^{2}A_{m}(Q^{2})L^{m},
\end{eqnarray}
describes fairly well the available experimental data on the
reduced cross sections, and it is  pertinent in investigations of
the scattering of cosmic neutrinos from hadrons. The effective
parameters read
\begin{eqnarray}
D(Q^{2})&=&\frac{Q^{2}(Q^{2}+\lambda
M^{2})}{(Q^{2}+M^{2})^{2}},\nonumber\\
A_{0}(Q^{2})&=&a_{00}+a_{01}L_{2}(Q^{2}),\nonumber\\
A_{i}(Q^{2})&=&\sum_{k=0}^{2}a_{ik}L_{2}(Q^{2})^{k},~~i=(1,2),\nonumber\\
L(Q^{2})&=&\ln{\frac{1}{x}}+{\ln}\frac{Q^{2}}{Q^{2}+\mu^{2}},\nonumber\\
L_{2}(Q^{2})&=&{\ln}\frac{Q^{2}+\mu^{2}}{\mu^{2}},
\end{eqnarray}
where, the coefficients are defined in Table I.\\
An analytical derivation of the gluon distribution function from
the known structure function $F^{p}_{2}(x,Q^2)$ (i.e., Eq.(15))
and its derivative $\frac{dF^{p}_{2}(x,Q^2)}{d{\ln}Q^2}$ to
include the effects of heavy-quark masses based on Laplace
transforms is extended in Refs.[25, 6]. The gluon distribution
function is derived directly using a Laplace transform method, for
four massless quarks, by the following form
\begin{eqnarray}
xg(x,Q^2)&=&\frac{9\pi}{5\alpha_{s}}\bigg{\{}3\mathcal{F}^{p}_{2}(x,Q^2)-x\frac{\partial}{\partial{x}}\mathcal{F}^{p}_{2}(x,Q^2)\nonumber\\
&&-\int_{x}^{1}\mathcal{F}^{p}_{2}(z,Q^2)\bigg{(}\frac{x}{z}\bigg{)}^{3/2}\bigg{[}
\frac{6}{\sqrt{7}}\sin(\frac{\sqrt{7}}{2}{\ln}\frac{z}{x})\nonumber\\
&&+2\cos(\frac{\sqrt{7}}{2}{\ln}\frac{z}{x})\bigg{]}\frac{dz}{z}\bigg{\}},
\end{eqnarray}
where
\begin{eqnarray}
\mathcal{F}^{p}_{2}(x,Q^2)&=&\frac{\partial{F}^{p}_{2}(x,Q^2)}{\partial{\ln}Q^2}-\frac{\alpha_{s}}{4\pi}\bigg{\{}
\int_{x}^{1}\frac{\partial{F}^{p}_{2}(z,Q^2)}{\partial{z}}{\times}\nonumber\\
&&\bigg{[}\frac{16}{3}{\ln}\frac{z}{z-x}-\frac{4}{3}\bigg{(}\frac{x^2}{z^2}+\frac{2x}{z}\bigg{)}\bigg{]}dz\bigg{\}}.
\end{eqnarray}
Here $\alpha_{s}$ is the running coupling at the LO approximation
with $\alpha_{s}(M_{z}^2)=0.118$ and
$\Lambda_{n_{f}=4}=120.4~\mathrm{MeV}$ and
$\Lambda_{n_{f}=5}=87.8~\mathrm{MeV}$. Similar investigations of
the longitudinal structure function, with respect to the Mellin
and Laplace transform methods, have been performed in
Refs.[8,10,26].\\
Parameterizations of the nuclear parton distribution functions
proposed by some groups such as Eskola, Kolhinen and Salgado (EKS)
[27], by de Florian and Sassot (DS) [28], by Hirai, S. Kumano and
T. H. Nagai (HKN) [29], by K. J. Eskola, H. Paukkunen and C. A.
Salgado (EPS) [30] and briefly discussed in Refs.[31-33]
respectively. Some recent works determined the nuclear partons in
Refs.[34-37]. In Ref.[34], the new nCETQ15 extends CTEQ proton
PDFs to include the nuclear dependence using data on nuclei all
the way up to $ ^{208}\mathrm{Pb}$ with uncertainties using the
Hessian method. An updated global analysis of collinearly
factorized nuclear parton distribution functions (nPDFs) at
next-to-leading order approximation (NLO) in perturbative QCD is
presented in Ref.[35] which includes more data from proton-lead
collisions at the Large Hadron Collider (LHC). The shadowing and
anti-shadowing effects on gluons in large nuclei are considered at
small and intermediate values of $x$. The uncertainties of nPDFs
are included within the Hessian framework. Global NLO QCD analysis
of hard processes in fixed-target lepton-nucleus and
proton-nucleus together with collider proton-nucleus experiments
are presented in Ref.[36] for determination of nuclear parton
distributions and provide predictions for ultra-high-energy
neutrino-nucleon cross-sections, relevant for data interpretation
at neutrino observatories. The extensive dataset underlying the
nNNPDF3.0 determines the shadowing of gluons and sea quarks as
well as the anti-shadowing of gluons at small and large $x$ values
respectively.\\
In the following, the HIJING [37,38] parametrization is
considered, which is in good agreement with the ALICE experiment
at LHC energies, which provides a more stringent constraint on
gluon shadowing. The nuclear modification factors in the HIJING
parametrization are given by
\begin{eqnarray}
R^{A}_{F_{2}}(x,b)&=& 1+1.19(\ln
A)^{1/6}(x^3-1.2x^2+0.21x)\nonumber\\
&&-s_{q}(b)(A^{1/3}-1)^{0.6}(1-3.5\sqrt{x})\nonumber\\
&&{\times}\exp(-x^2/0.01),
\end{eqnarray}
and
\begin{eqnarray}
R^{A}_{G}(x,b)&=& 1+1.19(\ln
A)^{1/6}(x^3-1.2x^2+0.21x)\nonumber\\
&&-s_{g}(b)(A^{1/3}-1)^{0.6}(1-1.5{x}^{0.35})\nonumber\\
&&{\times}\exp(-x^2/0.004),
\end{eqnarray}
where $s_{a}(b)$ reads
\begin{eqnarray}
s_{a}(b)=s_{a}\frac{5}{3}(1-b^2/R_{A}^{2}),
\end{eqnarray}
with  $s_{q} = 0.1$ and $s_{g}= 0.22-0.23$. Indeed, the impact
parameter dependence of the shadowing is implemented through the
parameters $s_{a}(b)$ in Eq.(21) of the HIJING2.0 model. The
authors, in Ref.[38], defined that the form of the impact
parameter dependence is chosen to give rise to the centrality
dependence of the pseudorapidity multiplicity density per
participant pair. Considering the dependence of the gluon density
on the impact parameter, shows that this is stronger than the
typical nuclear length $L_{A}=\sqrt{R_{A}^{2}-b^2}$ dependence.\\
For a nuclear target with the mass number A, we take
$R_{A}=1.25~\mathrm{fm}{\times}A^{1/3}$ which is the nuclear size.
The impact parameter of $b$ is chosen for the central with $b=0$
and the peripheral with $b=5~\mathrm{fm}$ (for heavy nuclei) [4].\\
Therefore, the nuclear structure functions can be defined and
parametrized via the HIJING and $F_{2}^{p}$ parametrizations as we
find that
\begin{eqnarray}
F_{2}^{A}(x,Q^2,b)&=&AF_{2}^{p}(x,Q^2)(\mathrm{i.e.,Eq.15})\nonumber\\
&&{\times}R^{A}_{F_{2}}(x,b)(\mathrm{i.e.,Eq.19}),
\end{eqnarray}
and
\begin{eqnarray}
F_{L}^{A}(x,Q^2,b)&=&A\frac{10\alpha_{s}}{27\pi}xg^{p}(kx,Q^2)(\mathrm{i.e.,Eq.17})\nonumber\\
&&{\times}R^{A}_{G}(kx,b)(\mathrm{i.e.,Eq.20}).
\end{eqnarray}
So, the nuclear reduced cross section is parametrized by the
following form
\begin{eqnarray}
\sigma_{r}^{A}(x,Q^2,b)&=&AF_{2}^{p}(x,Q^2)R^{A}_{F_{2}}(x,b)-\frac{y^2}{Y_{+}}A\frac{10\alpha_{s}}{27\pi}\nonumber\\
&&{\times}xg^{p}(kx,Q^2) R^{A}_{G}(kx,b).
\end{eqnarray}
In the following, the nuclear structure functions, $F_{2}^{A},
F_{L}^{A}$ and the nuclear reduced cross section,
$\sigma_{r}^{A}$, are determined owing to the EIC data range.\\

\subsection{IV. Results and Conclusions}

The kinematic regions at the EIC are proposed with
$\sqrt{s}=140~\mathrm{GeV}$ where the numerical results are
determined by inelasticity $y{\lesssim}1$. The impact parameter
$b$ is selected  to be $b=0$ and $5~\mathrm{fm}$ for heavy nuclei
(i.e., Pb-208) and $b=0$ for light nuclei (i.e., C-12). The gluon
density is expanded when the points $a=0$ and $0.9$ are used. The
results for the ratios $R_{F_{L}}^{A}$ and $R_{\sigma_{r}}^{A}$
and the nuclear structure functions, $F_{2}^{A}$ and $F_{L}^{A}$
and the nuclear reduced cross section $\sigma_{r}^{A}$ are
obtained at scales $Q^2=5~\mathrm{GeV}^{2}$ and
$10~\mathrm{GeV}^{2}$ for the nucleus of Pb-208 and C-12 in
Figs.1-4 respectively. The ratio of gluon and singlet
distributions are used from the HIJING parameterization. In these
figures (in the left panels), the ratios $R_{F_{2}}^{A}$ (HIJING
parametrization, Eq.(19) ), $R_{F_{L}}^{A}$ (Eq.(13)) and
$R_{\sigma_{r}}^{A}$ (Eq.(14)) are shown for lead (Pb-208) at
$Q^2=5~\mathrm{GeV}^2$ (Fig.1) and $Q^2=10~\mathrm{GeV}^2$ (Fig.2)
and for carbon (C-12) at $Q^2=5~\mathrm{GeV}^2$ (Fig.3) and
$Q^2=10~\mathrm{GeV}^2$ (Fig.4) respectively. In the middle
panels, the structure functions $F_{2}^{A}$ (Eq.(22) ),
$F_{L}^{A}$ (Eq.(23)) and the reduced cross section
$\sigma_{r}^{A}$ (Eq.(24)) are shown for lead (Pb-208) carbon
(C-12) at $Q^2=5~\mathrm{GeV}^2$ and $Q^2=10~\mathrm{GeV}^2$ in
Figs.1-4 respectively. In the right panels, the functions divided
by A for the heavy nucleus of Pb-208 and for the light nucleus of
C-12 as a function of the momentum fraction $x$ are shown.\\
We observe that the results are dependent on the impact parameter
and the expansion point. The results in these figures (i.e.,
Figs.1-4) in the first row are dependent on the impact parameter,
and they are independent of the expansion point. But in the two
and three rows, the results are dependent on those. In the first
row, the results increase as the impact parameter increases. The
results for the heavy nucleus of Pb-208 are sensitive for
different values of $b=0$ and $5~\mathrm{fm}$ and for the light
nucleus of C-12 are dependent on the  value of $b=0$. In the
second row, the ratio $R^{A}_{F_{L}}$ is dependent on the
coefficients $a$ and $b$. The results for $F_{L}^{A}$ and
$F_{L}^{A}/A$ increase as the impact parameter increases and
decreases as the expanding point increases. This is visible as the
longitudinal structure function at $x$ is proportional to the
gluon density at $kx$. These upper and lower values are comparable
with the original NNPDF3.1 PDF uncertainties [39] as considered in
Ref.[40].\\
In the third row, the ratio $R^{A}_{\sigma_{r}}$ is independent of
the coefficient $a$ at $b=0$ and is dependent on it at
$b=5~\mathrm{fm}$ for $x{\leq}10^{-3}$. The increase of
$R^{A}_{\sigma_{r}}$ towards low $x$, reflecting the fall of the
reduced cross sections (nucleon and nuclei) towards larger $y$
region. This behavior is observable for $\sigma^{A}_{r}$ and
$\sigma^{A}_{r}/A$ at low $x$ with increases of  the impact
parameter and the expansion point for the heavy nucleus of Pb-208
and is visible at $b=0$ for the light nucleus of C-12. The results
increase as the $Q^2$ values increase in Figs.2 and 4. The
importance of the nuclear longitudinal structure function
$F_{L}^{A}$ at low $x$ in a wide range of the mass number $A$ in
all figures is observable as the contribution of $F_{L}^{A}$ is
enhanced with $y^2$ and the nuclear reduced cross section
$\sigma^{A}_{r}$ tends
to $F_{2}^{A}-F_{L}^{A}$ for $y{\rightarrow}1$.\\

In conclusion, we have discussed the determination of the nuclear
reduced cross section and its ratio in the EIC kinematic range for
the heavy and light nucleus Pb-208 and C-12 respectively. The
HIJING parametrizations for $R^{A}_{F_{2}}$ and $R^{A}_{G}$ is
used for the nuclear distributions. The nuclear longitudinal DIS
structure function $F_{L}^{A}(x,Q^2)$ at small $x$ is obtained in
terms of the effective parameters of the Froissart-bound
parametrization of $F_{2}(x,Q^2)$ and $R^{A}_{F_{2}}$ owing to the
expanding point of the gluon density. In the EIC kinematic range,
the importance of the longitudinal structure function for nuclei
should be required, since the nuclear reduced cross section is
dependent on it at high inelasticity. The $x$ dependence of the
nuclear longitudinal structure function and nuclear reduced cross
section  at $Q^2=5$ and $10~\mathrm{GeV}^2$ in a wide range of
nuclei have been calculated. The magnitude of these results
increases with a decrease of $x$ and an increase of the atomic
number A and $Q^2$. The results are dependence on the impact
parameter of the light and heavy nuclei and the expanding point of
the gluon density at small $x$. The nuclear reduced cross section
divided by A has a depletion with increases of $y$ similar to the
reduced cross section of the proton in the EIC kinematic range and
its dependent on the impact parameter and the expanding point. We
hope that this paper can motivate a more accurate determination of
 $F_{L}^{A}$ and $\sigma_{r}^{A}$ in the next years in the EIC
 colliders.\\

\subsection{ACKNOWLEDGMENTS}

The authors are thankful to the Razi University for financial
support of this project.

\begin{table}[h]
\caption{ The coefficients [5] at low $x$ for
$0.15~\mathrm{GeV}^{2}<Q^{2}<3000~\mathrm{GeV}^{2}$.}
\begin{tabular} {cccc}
\toprule \\  \multicolumn{2}{c}{parameters \quad \quad \quad ~~~~~~~~~~~~~~~~value}    \\ &&&\\ \hline \\ &&&\\

$a_{00}$ & \quad $ 0.255 \pm 0.016$ & &\\

$a_{01}$& \quad  $1.475\times 10^{-1}~\pm 3.025\times10^{-2}$ & &\\&&&\\

  $a_{10} $  &   \quad  $8.205\times 10^{-4}~~  \pm  4.62\times10^{-4} $  \\

  $a_{11} $  &   \quad   $-5.148\times 10^{-2}\pm 8.19\times10^{-3}$  \\

  $a_{12}$   &    \quad  $-4.725\times 10^{-3}\pm 1.01\times10^{-3}$   \\  &&&\\

 $a_{20}$   &   \quad   $2.217\times 10^{-3}\pm 1.42\times10^{-4} $ \\

 $a_{21}$   &   \quad   $1.244\times 10^{-2}\pm 8.56\times10^{-4}$  \\

 $a_{22}$    &    \quad  $5.958\times 10^{-4}\pm 2.32\times10^{-4} $ \\ &&& \\

$n$& \quad  $11.49\pm 0.99$ & &\\

$\lambda$& \quad  $2.430~\pm 0.153$ & &\\

$M^{2}$ & \quad $0.753 \pm 0.068~ \mathrm{GeV}^{2}$ & &\\

$\mu^2$ & \quad $ 2.82 \pm 0.290~ \mathrm{GeV}^{2}$ & &\\

$\chi^{2}(\mathrm{goodness~ of~ fit})$ &  \quad  $0.95$ & &\\
\hline

\end{tabular}
\end{table}


\newpage
\section{References}
1. J. Aubert et al. (European Muon Collaboration), Phys. Lett.B
{\bf123}, 275 (1983).\\
2. L.S.Moriggi, G.M.Peccini and M.V.T.Machado,
Phys.Rev.D {\bf103}, 034025 (2021).\\
3. Anna M.Stasto, Acta Physica Polonica B {\bf16}, 7-{\bf A}23
(2023).\\
4. N.Armesto, Eur.Phys.J.C {\bf26}, 35 (2002).\\
5. M. M. Block, L. Durand and P. Ha, Phys.Rev.D {\bf89}, 094027 (2014).\\
6. Martin M. Block, Loyal Durand and Douglas W. McKay, Phys.Rev.D
{\bf79}, 014031 (2009).\\
7. M. Froissart, Phys. Rev. {\bf123}, 1053 (1961).\\
8. L.P. Kaptari, et al., JETP Lett. {\bf109}, 281 (2019).\\
9. G.R. Boroun, Phys.Rev.C {\bf97},  015206 (2018).\\
10. L.P. Kaptari, et al., Phys.Rev.D {\bf99},  096019 (2019).\\
11. G.R.Boroun and B.Rezaei, Phys.Lett.B {\bf816}, 136274 (2021).\\
12. K.J. Eskola, H. Honkanen, V.J. Kolhinen and C.A. Salgado,
Phys.Lett.B {\bf532}, 222 (2002).\\
13. Jamal Jalilian-Marian and Xin-Nian Wang, Phys.Rev. D {\bf63},
096001 (2001).\\
14. N. Armesto, J. Phys. G {\bf32}, R367 (2006).\\
15. F. Gelis, E. Iancu, J. Jalilian-Marian, and R. Venugopalan,
Annu. Rev. Nucl. Part. Sci. {\bf60}, 463 (2010).\\
16. R. Abdul Khalek et al., Snowmass 2021 White Paper, arXiv
[hep-ph]:2203.13199.\\
17. R.Abir et al., The case for an EIC Theory Alliance:
Theoretical Challenges of the EIC, arXiv [hep-ph]:2305.14572.\\
18. LHeC Collaboration, FCC-he Study Group, P. Agostini, et al.,
J. Phys. G, Nucl. Part. Phys. {\bf48},  110501 (2021).\\
19. E.Sichtermann, Nucl.Phys.A {\bf956}, 233 (2016).\\
20. E. C. Aschenauer et al., Phys. Rev. D {\bf96}, 114005
(2017).\\
21. M. Ericson and S. Kumano, Phys.Rev. C {\bf67},  022201
(2003).\\
22. G. Altarelli and G. Martinelli, Phys. Lett. B {\bf76}, 89 (1978).\\
23. M.B.Gay Ducati and P.B.Goncalves, Phys.Lett.B {\bf390}, 401
(1997).\\
24. G.R.Boroun and B.Rezaei, Eur.Phys.J.C 72, 2221 (2012).\\
25. M. M. Block and L. Durand, arXiv[hep-ph]:0902.0372.\\
26.  G.R.Boroun and B.Rezaei, Phys.Rev.D {\bf105}, 034002 (2022);
G.R.Boroun and B.Rezaei, Phys.Rev.C {\bf107}, 025209
(2023).\\
27. K. J. Eskola, V. J. Kolhinen and C. A. Salgado, Eur. Phys. J.
C {\bf9}, 61 (1999).\\
28. D. de Florian and R. Sassot, Phys. Rev. D {\bf69}, 074028
(2004).\\
29. M. Hirai, S. Kumano and T. H. Nagai, Phys. Rev. C {\bf76},
065207 (2007).\\
30. K. J. Eskola, H. Paukkunen and C. A. Salgado,
arXiv[hep-ph]:0802.0139.\\
31. N. Armesto, J. Phys. G {\bf32}, {R}367 (2006).\\
32. E.R. Cazaroto, F. Carvalho, V.P. Goncalves and F.S. Navarra,
Phys.Lett.B {\bf669}, 331 (2008).\\
33. Wei-tian Deng, Xin-Nian Wang and R. Xu, Phys.Lett.B {\bf701},
133 (2011).\\
34. [nCTEQ15 Collaboration] K.Kovarik et al., Phys. Rev. D
{\bf93},
085037 (2016).\\
35. [nNNPDF3.0 Collaboration] R.Abdul Khalek et al., Eur. Phys. J.
C {\bf82}, 507 (2022).\\
36. [EPPS21 Collaboration] K. J. Eskola, P. Paakkinen, H.
Paukkunen and C. A. Salgado,
Eur. Phys. J. C {\bf82}, 413 (2022).\\
37. S. -Y. Li and X. -N. Wang, Phys. Lett. B {\bf527}, 85
(2002).\\
38. W. -T. Deng, X. -N.Wang and R. Xu, Phys.Rev.C {\bf83}, 014915
(2011).\\
39. NNPDF collaboration, R. D. Ball et al., Nucl. Phys. B
{\bf849}, 112 (2011).\\
40. N.Armesto et al., Phys. Rev. D {\bf105}, 114017 (2023).\\


\begin{figure}
\includegraphics[width=1\textwidth]{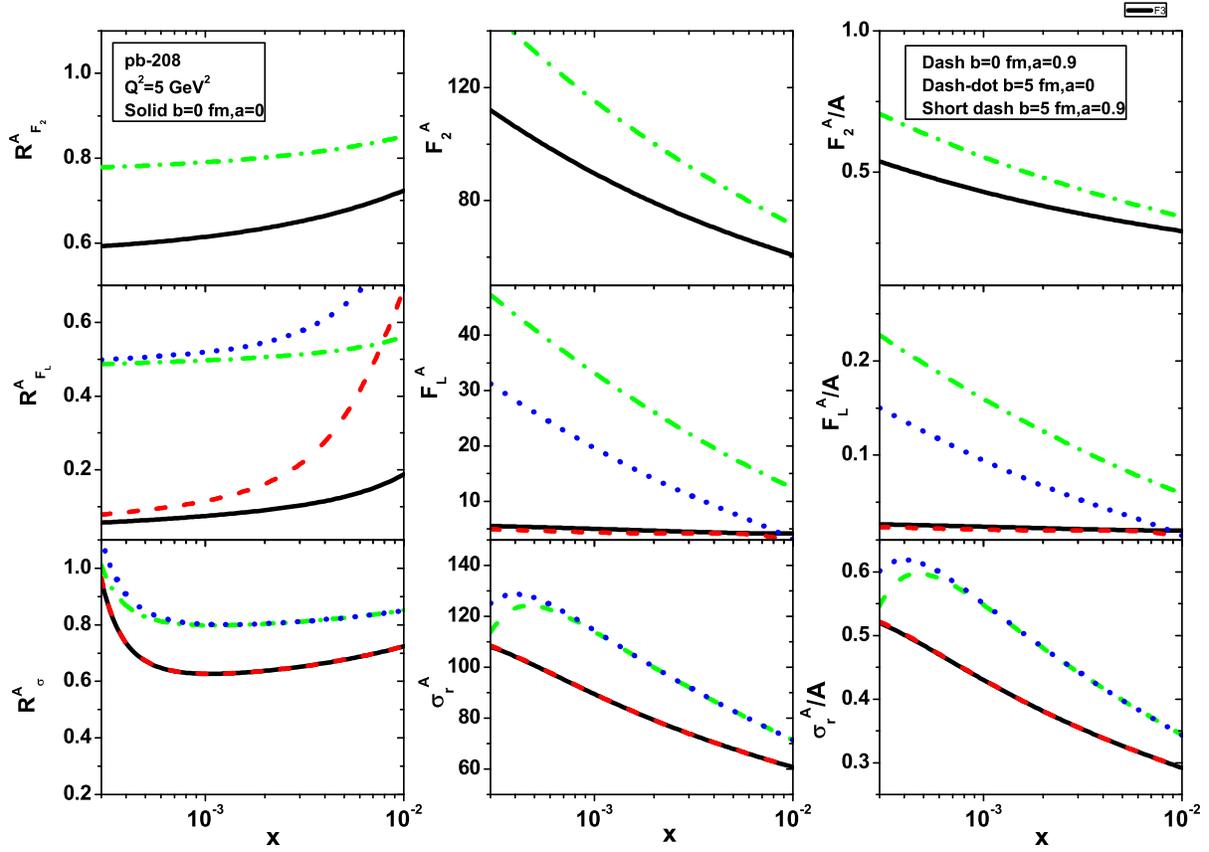}
\caption{Results of the ratios, structure functions and reduced
cross section of the nucleus of Pb-208 are shown as a function of
$x$ at $Q^2=5~\mathrm{GeV}^2$. We observe, in the left panel, the
ratios $R_{F_{2}}^{A}$, $R_{F_{L}}^{A}$ and $R_{\sigma_{r}}^{A}$;
in the middle panel, $F_{2}^{A}$, $F_{L}^{A}$ and $\sigma_{r}^{A}$
; in the right panel, $F_{2}^{A}/A$, $F_{L}^{A}/A$ and
$\sigma_{r}^{A}/A$ from up to down respectively. The curves show
the results owing to the impact parameters and the expanding
points as: $b=0~\mathrm{fm}$ and $a=0$ (solid black),
$b=0~\mathrm{fm}$ and $a=0.9$ (dash red), $b=5~\mathrm{fm}$ and
$a=0$ (dash-dot green) and $b=5~\mathrm{fm}$ and $a=0.9$ (short
dash blue) respectively.}\label{Fig1}
\end{figure}

\begin{figure}
\includegraphics[width=1\textwidth]{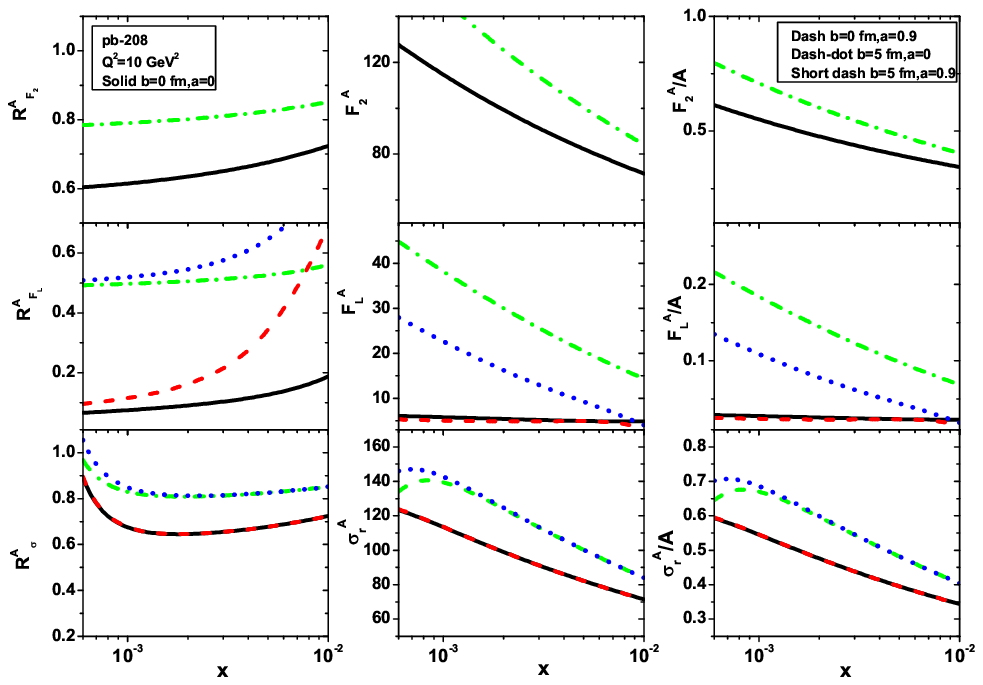}
\caption{The same as Fig.1 for the nucleus of Pb-208  at
$Q^2=10~\mathrm{GeV}^2$. }\label{Fig2}
\end{figure}

\begin{figure}
\includegraphics[width=1\textwidth]{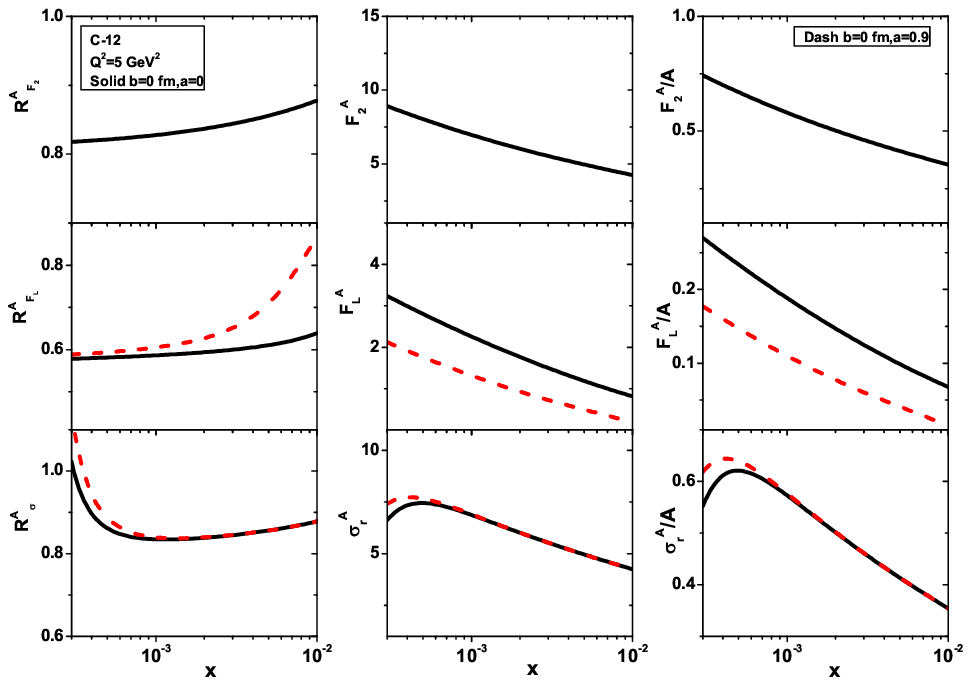}
\caption{The same as Fig.1 for the nucleus of C-12  at
$Q^2=5~\mathrm{GeV}^2$.}\label{Fig3}
\end{figure}

\begin{figure}
\includegraphics[width=1\textwidth]{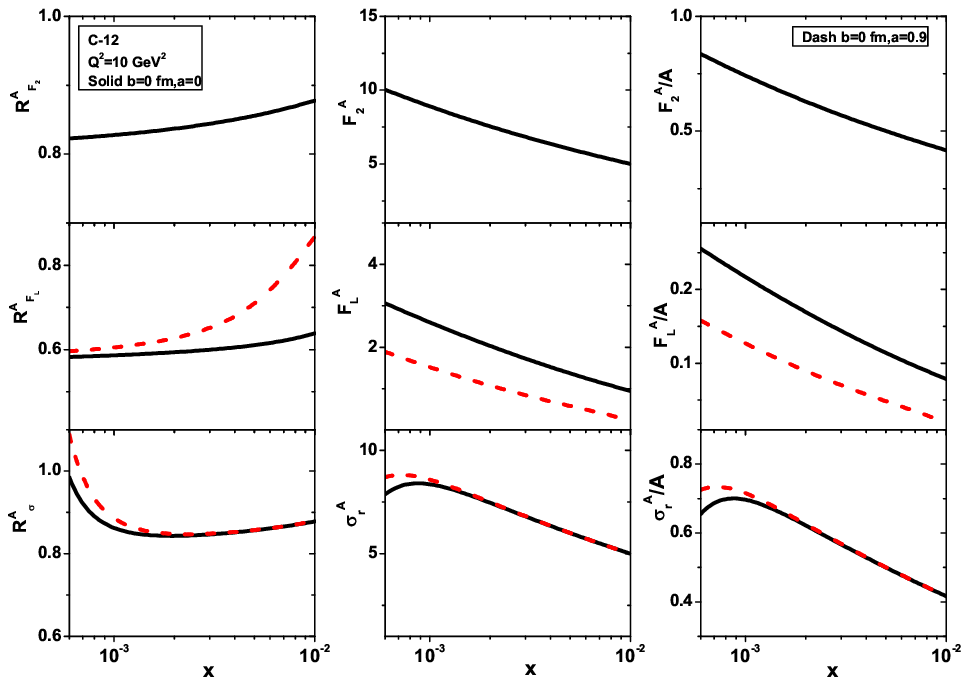}
\caption{The same as Fig.1 for the nucleus of C-12  at
$Q^2=10~\mathrm{GeV}^2$.}\label{Fig4 }
\end{figure}


\end{document}